# Thermostatistics of a single particle on a granular dimer lattice: influence of defects

K. Combs and J. S. Olafsen

Department of Physics, One Bear Place, #97316, Baylor University, Waco, Texas 76798,

USA

A. Burdeau and P. Viot

Laboratoire de Physique Theorique de la Matiere Condensee, UMR 7600, Universite

Pierre et Marie Curie, 4, place Jussieu 75752 Paris Cedex 05, France

We study the thermostatistical fluctuations of a single Delrin monomer on a granular lattice of dimer particles using both experiment and simulation. The goal is to examine the collision frequency, energy injection, and sidewall effects on a single second-layer particle in a bi-layer granular gas experiment. Non-Gaussian velocity statistics are observed for the single particle of the top layer and result from the presence of defects in the first layer. These deviations are not directly due to the presence of the boundary wall, since the form of velocity distributions is quite spatially homogeneous, but are the consequence of the presence of a few mobile defects in the first layer.

Recent measurements in a bi-layer granular experiment have shown that under certain conditions, robust Gaussian velocity statistics may be obtained for a wide variety of shaking parameters in the dynamics of a homogenous layer of identical monomer spheres in the upper layer [1,2,3]. Interestingly enough, in the same experiment, the velocity statistics for the lower layer in the mechanically driven experiment simultaneously exhibit strongly non-Gaussian behavior [1,2,3,4]. The coincidental behavior is all the more interesting since the two layers are in "thermal contact" with each other, exchanging energy through collisions between particles within the same layer (intralayer) and particles from different layers (interlayer). This dichotic process is suggested to satisfy the Central Limit Theorem and explain the robust Gaussian statistics observed in both the experiment and a recent simulation[5]. In a system driven far from equilibrium, it has been noted that a statistical mechanics approach can be regained[6], but only if certain conditions are met[7].

Both layers are driven far from equilibrium via mechanical shaking, yet kept segregated through certain physical aspects of the constituent particles. In this brief report, we probe the mechanism for this robust behavior in the velocity probability distribution functions (pdfs) by examining the velocity statistics for a single second-layer Delrin particle free to move on top of a granular dimer lattice which, in turn, is driven by a mechanically shaken plate. Additionally, the cell geometry is chosen for these measurements to also aid in determining the relative effect of the sidewalls and the system size.

Whereas the initial geometry was a horizontal flat plate of diameter, D = 29.2 cm, that was mechanically shaken in the vertical direction to drive the motion of 3,288 dimers comprised of two spheres each of diameter d = 3.2 mm and loosely connected by a metal rod that allowed the two spheres to separate up to 1.6 mm, the cell for the measurements here has a diameter of 2.54 cm upon which 23 horizontally oriented dimers are driven. In the prior experimental work[2], several thousand upper layer particles of either diameter 3.2 mm or 4.76 mm were used. Here, a single Delrin monomer of diameter 4.76 mm is the sole occupant of the upper layer in the cell. The larger Delrin species was used in this investigation due to the propensity of the smaller Delrin tracer to become trapped within the lower layer defects.

The motion of both the lower layer dimers and upper layer monomer are tracked using the same Dalsa CA-D6 260x260 pixel, 955 fps camera as was used in the larger cell measurements[1,2,3,4]. Similarly, particles in the lower dimer layer and the upper layer Delrin particle are then tracked using the same software algorithms as previously written to collect ensemble measurements of the velocity statistics in the two layers. As there is only a single upper layer particle in these measurements, a larger set of ensemble measurements were necessary to generate sufficiently large statistics. Measurements reported here were obtained at a shaking frequency of f = 70 Hz and a dimensionless root mean square (rms) acceleration, $\Gamma_{rms} = A_{rms} (2 \pi f)^2/g = 2$, where g is the acceleration due to gravity and $A_{rms}$ is the rms amplitude of the plate shaking. In the prior work for the larger cell[1,2,3], measurements had been obtained at shaking frequencies of 50, 70 and 90 Hz at $\Gamma$ = 2.25, 2.0 and 1.75 for second layer coverages of 20, 40, 60 and 80%.

Physically, the only differences between this cell and the previous experiments on bilayer systems[1,2,3] are (a) the smaller cell diameter, (b) the associated smaller number of dimers comprising the first layer, and (c) the use of a single, larger diameter Delrin particle in the second layer instead of a large number. Each of these differences allow for determining the contribution of (a) the effect of the sidewalls; (b) the number of particles in the lower dimer layer; and (c) the collisions within the upper layer to the robust Gaussian and non-Gaussian velocity statistics observed in the upper and lower layers of the larger experiment, respectively. A sample image from the smaller cell configuration is shown in Figure 1. A total of ~135,000 frames were analyzed for the data in this paper to achieve statistics on the same order as those from prior investigations of the dynamics in the larger version of this bilayer experiment that had many upper layer particles.

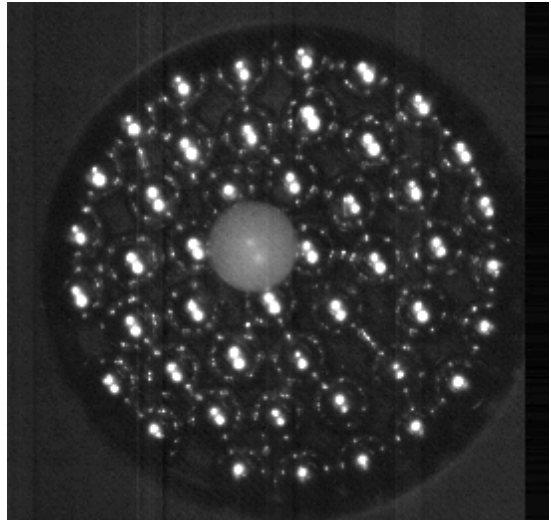

**Figure 1: A view from above of the experimental cell. A single Delrin particle moves on top of a layer of stainless steel dimers. Note the defects currently present in the lower left and right areas of the dimer layer.**

To determine the effects of system size and the sidewalls, velocity data obtained in each layer is analyzed for regions of different radii as measured from the center of the cell. Figure 2 demonstrates how data is segregated for the motion of the single Delrin particle. To look for spatial dependencies of the rms velocities and the flatness of the distributions, the velocity pdfs were examined for data within areas of different radii from the center of the cell. In both the upper and lower layers, radii, r, of 45, 65, 75, 85 and 95 pixels were used. The interior wall of the cell was located at a radius $R = 110$ pixels from the center and the finite size of the particles precluded locations within a particle radius of the wall. The sorting was cumulative: the data for r = 95 pixels encompassed all the motion within the experimental cell (in either layer), etc.

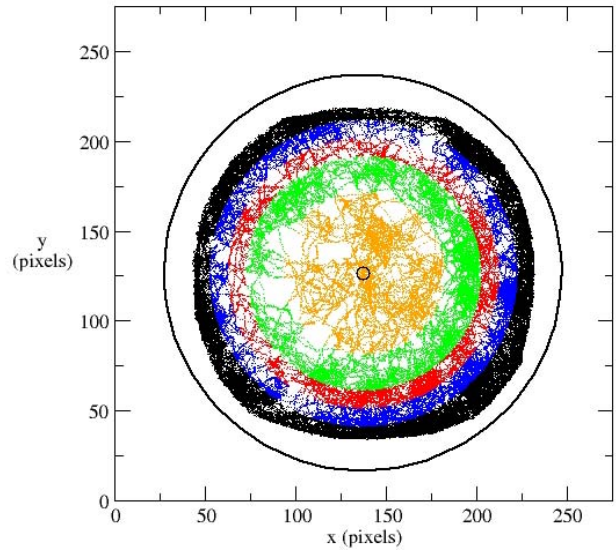

**Figure 2: (Color online) Upper layer particle tracks in the experimental cell. The large circular black line is the interior edge of the cell at R = 110 pixels from the center. Colored regions correspond to radial cuts made in data analysis. Colors are orange, r/R = 0.409; green, r/R = 0.591; red, r/R = 0.682; blue, r/R = 0.773; and black, r/R = 0.864.**

The velocity pdfs for the corresponding spatial cuts of the data in the lower layer are shown in Figure 3. All of the distributions show the same significant deviations from Gaussian (dashed line) behavior. Simulation of this system has been done with a discrete element method by using viscoelastic spheres. The particle-particle interaction is represented by a spring-dashpot model which consists of a linear elastic force and a viscous dissipation in the normal direction to the contact. To have a simple and realistic system a tangential force related to the normal force and respecting Coulomb friction is added. This simple model allows us to easily tune the dissipation in the system via constant coefficients of restitution. The values of these parameters have been chosen so that the numerical temperature corresponds to the experimental measure. The coefficient of restitution was set to 0.3 for the collisions within the first layer, to account for the dissipative nature of the dimers, to 0.5 for the collisions of the dimers with the bottom cell, and to 0.7 for the collisions between the tracer and the dimers. The geometry of the simulation cell corresponds to that of the experimental cell with the same dimensions. The cell bottom is vibrated sinusoidally as in the experiment. However, the first-layer composed of dimers in experiments is made of spherical monomers in simulation. This simplification is irrelevant for close packed layers but the coverage in the small cell studied here is inferior to the close packing coverage. (The boundaries prevent a perfect hexagonal packing and induce defects – see the lower right region of Fig. 1.)

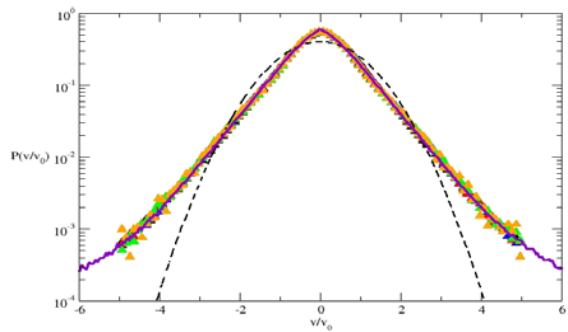

**Figure 3: (Color online) Velocity pdfs for the lower dimer layer. Solid line is the simulation and dashed line is a Gaussian function. Colors are orange, r/R = 0.409; green, r/R = 0.591; red, r/R = 0.682; blue, r/R = 0.773; and black, r/R = 0.864.**

The coverage of the first layer is equal to 0.73 in the experimental setup, and by using spheres (instead of dimers) the probability of a fluctuation allowing room for the tracer particle to fall into the monomer layer is high, and if that were allowed to occur, the single particle of the top layer would become trapped during the simulation run. To avoid this behavior that the dimer bars suppress, the simulation was modified to account for the existence of dimer connectors – at least in a mean-field manner: when a suitably large enough hole appears, i.e. the horizontal height of the center of the tracer particle is less than 0.93 diameters above the plate, a collision between the top particle and a virtual floor occurs. This ad hoc perturbation maintains the separation of the tracer from the lower layer and models a "collision" with a virtual bar that should be connecting the monomers as dimers. No modification is made for the dynamics of the first layer particles and only the tracer adsorption into the first layer is prevented by incorporating this simple rule. As the floor does not reproduce details concerning the rigidity of the dimer; a slightly higher density for the lower layer was initially used: the simulation first consisted of 52 particles versus 23 dimers in the experiments, corresponding to a coverage c=0.82. The exact corresponding number of monomers, namely 46, has also been simulated, showing a good agreement with the experimental results and indicating an increased tendency of the defect-related phenomenon with the decrease of the lower layer density.

The interesting consequential behavior is observed in the velocity distribution for the upper layer single Delrin particle. The distributions for discriminating the data within different radii from the center of the cell are shown in Figure 4. While there is no apparent trend with regard to the size of the spatial cut across the cell, the tails of the single particle pdfs and the simulation previously created for the bilayer experiments for much larger numbers of upper layer particles (solid line) nonetheless demonstrate significant deviations from a Gaussian (dashed line). The flatness, a ratio of the fourth moment of the distribution to the square of the second moment, is a sensitive measure of the tails of the distribution. Recall that the flatness is equal to 3 when the distribution is Gaussian.

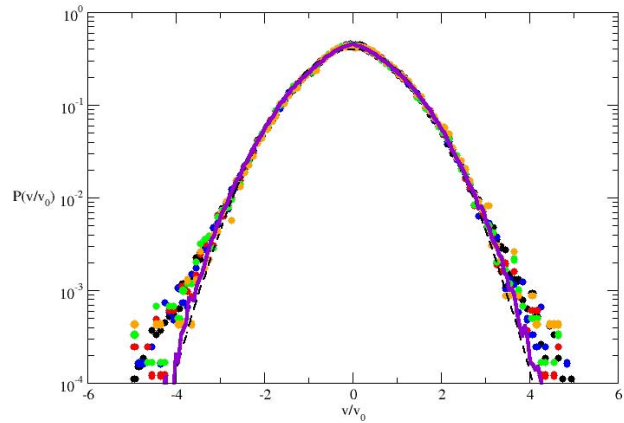

**Figure 4: (Color online) Velocity pdfs for the single upper layer particle. Solid line: simulation, Dashed line: Gaussian function. Colors are orange, r/R = 0.409; green, r/R = 0.591; red, r/R = 0.682; blue, r/R = 0.773; and black, r/R = 0.864.**

Figure 5 is a plot of the flatness, F, of the bottom (triangles) and upper (circle) particle velocity pdfs in the experiment. The measure of the flatness in each of the two layers does not demonstrate any strong spatial dependences within the smaller cell, however, there is a slight upturn in the flatness for data at the wall (black symbols, the cut corresponding to r = 95 pixels) that may be indicative of a slight increase in the dissipation at the lateral edge of the cell or the tendency of the defects in the dimer layer to appear at the wall. The overall average value of both the upper and lower layer flatness is larger than that measured in the larger diameter cell[2]. In addition, the general trend in the upper layer data from the larger cell implied a tendency for the flatness to actually decrease (become more Gaussian) as the coverage of the upper layer was decreased from 80% to 20%. For most of the data in the larger cell, the flatness in the upper layer was on the order of 3.01 – 3.1, denoting a relatively robust Gaussian distribution. However, in the case of a single upper layer particle, the deviation of the average flatness from a Gaussian is significant at a value of F = 3.42. To understand this behavior, we return to our discussion of the virtual floor used to model the dimer connecting rods in a mean field manner.

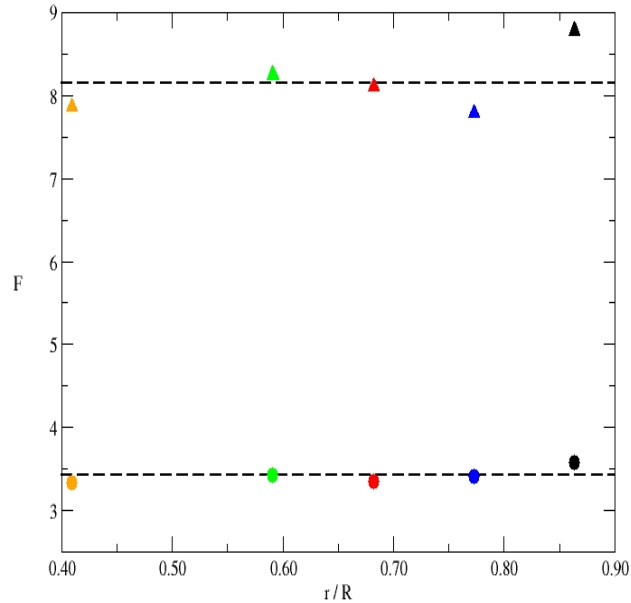

**Figure 5: Flatness of the velocity pdfs in the experiment: lower (triangles) and upper (circles) layers. Bottom Layer Average Flatness: 8.15, Top Layer Average Flatness: 3.42**

The virtual floor above the vibrating plate can be assumed either elastic or inelastic (to quantify the role of inelasticity of the rod). The velocity distributions for the upper tracer particle obtained in both of these two situations do not exhibit quantitative differences, indicating that dissipation due to the presence of the floor/dimer rod does not play a central role in the results obtained. The flatness of the distributions obtained in both of these simulations at $\Gamma = 2$, $f = 70 Hz$ were F = 3.2 +-1% when sampled in a central disc of half the radius of the cell, and F=3.25 +-1% when sampled in the full cell, for the experimental coverage N = 46 particles in the lower layer. The simulation with a denser coverage for the lower layer (N = 52 particles) gave F=3.2+-1% in the full cell as well as in the central half disc. In the first case the defects play a more significant role in the dynamics of the tracer due to the very loose character of the lower layer. For comparison, simulation of the tracer on an initially hexagonally packed monomer lower layer in a rectangular box with periodic boundary conditions has also been performed. In this case, the lower layer particles maintain nearly perfect hexagonal packing and the velocity distributions for the tracer are closer to Gaussian, with F = 3+-1%. This clearly shows that the lower layer packing defects are responsible for the deviation from the Gaussian. These defects are more numerous near the sidewall, which explains the greater deviations observed in simulations and the experiment when the velocities are sampled in the full cell.

In bilayer experiments[2] in a large cell, when increasing the coverage of the top layer, deviations from Gaussian increase. Similar trends are also verified by simulation results in the larger cell bilayer system which show a continuous increase in the flatness with the coverage of the upper layer, namely from F = 3 when c=0.2 to F=3.25 when c=0.8, in a rectangular periodic system. Collisions between particles of the same layer are expected to increase the flatness of the velocity distribution. Conversely, by using a single Delrin particle, a quasi-Gaussian velocity distribution was expected.

However, in the simulation, if the bottom layer is frozen as a fixed hexagonal or square lattice that does not fluctuate and only moves with the shaking plate[8], the system exhibits significant differences with the cases discussed above, the flatness observed for the velocity distributions being of order 2.9+-1%. And as previously explained, the distribution in the case of the tracer on an unfrozen lower layer shows F=3 in the absence of defects. The present simulation shows that the first layer coverage via its defect density induces deviations from Gaussian to the velocity distributions and explains the phenomenon observed in the small cell. Additional simulations in the small cell where the first layer coverage is increased (and consequently the defect density decreases) exhibit a decrease of the flatness.

The velocity distribution remains very homogeneous (See Figs. 3, 4 & 5), which implies defects are not localized in the cell, but are mobile during the experiment, and at any position of the cell an environment of the first layer evolves with time, but appears the same on average. The reason why the defects modify the velocity distribution is probably due to the fact that, when a particle "falls" into a defect, a transient "rattling effect" is introduced until the particle leaves this trap and restarts a diffusion process over the first layer. This multiple collision phenomenon tends to overpopulate the small velocity distribution. A very dense first layer prevents this effect. In the larger cell bi-layer granular gas experiments[1, 2, 3], the lower layer defects were "stirred" or annealed to the edges of the circular cell at the beginning of the data acquisition. Random defects in the dimer layer that diffused out through the experimental cell over time were likely responsible for the lack of a definite trend in the ensemble statistics that were obtained. Additionally, the intralayer Delrin collisions in the larger cell suppress the "trapping" and transient interlayer collisions that result in the non-Gaussian tails in the smaller cell/single Delrin particle presented here. The intralayer Delrin collisions are therefore even more important when the upper layer monomers are smaller than the tracer used here.

Authors (KC, JSO) acknowledge the support in part by funds from the Vice Provost for Research at Baylor University.

[1] G. W. Baxter and J. S. Olafsen, Nature (London) **425**, 680 (2003).

[2] G. W. Baxter and J. S. Olafsen, Granular Matter **9**, 135 (2007).

[3] G. W. Baxter and J. S. Olafsen, Phys. Rev. Lett. **99**, 028001 (2007).

[4] J. Atwell and J. S. Olafsen, Phys. Rev. E **71**, 062301 (2005).

[5] A. Burdeau and P. Viot, arXiv:0710.3713.

[6] D. A. Egolf, Science **287**, 101 (2000).

[7] R. P. Ojha, P.-A. Lemieux, P. K. Dixon, A. J. Liu, and D. J. Durian, Nature (London) **427**, 521 (2004).

[8] A. Prevost, D. A. Egolf, and J. S. Urbach, Phys. Rev. Lett. **89**, 084301 (2002).